# Acustica con una Bic e uno smartphone

L. Galante[#, 1], A. M. Lombardi[*, 2]
[#] LSS "G. Bruno", Torino, Italy.
[*] ITCS "Primo Levi", Bollate (MI), Italy.

**Abstract.** A smartphone, with its integrated sensors and cpu, can aid experiments in many different areas of Physics. We show how the resonant frequencies of a pipe can be measured using a smartphone and a Bic pen.

## I. INTRODUZIONE

Lo smartphone è considerato un nemico dell'apprendimento. Eppure esso offre enormi possibilità per adottare un approccio alla didattica che stimoli un maggiore coinvolgimento dei ragazzi e un uso di abilità intellettuali di più alto livello. Da qualche tempo è nata una piccola online-community, CLL (Comoving Little Lab, https://sites.google.com/site/comovinglittlelab), che conduce ricerche sull'utilizzo dei sensori interni ai cellulari per condurre esperimenti di fisica con ragazzi tra i 13 e i 19 anni. Questo articolo è un esempio del lavoro prodotto.

## II. ACUSTICA DI UNA PENNA BIC

L'insegnante di fisica sta spiegando le principali proprietà delle onde meccaniche e sente avvicinarsi con disagio lo scoglio di un argomento che lo obbligherà a qualche riga di calcoli: le onde stazionarie. Gli studenti guardano la lavagna riempirsi di formule e simboli, con un occhio. Con l'altro, uno sguardo al cellulare... quanto manca alla fine dell'ora? Ad un tratto il docente dice: "Ok ragazzi, adesso prendete i vostri cellulari ed una biro". La classe si anima di colpo. Come è possibile? I cellulari, nascosti nei posti più impensabili, ora devono uscire allo scoperto? Ed è l'insegnante a chiederlo! Sì, è proprio così. Basta un po' di fantasia e la classe può diventare un laboratorio di fisica, impensabilmente aggiornato e al passo coi tempi.

Negli astucci troviamo facilmente delle penne Bic. Togliamo il refill e la penna Bic è diventata un magnifico tubo risonante (non molto diverso da una canna d'organo). Il soffio dei ragazzi sull'apertura del

1   e.mail: lorenzo_galante@yahoo.it
2   e.mail: anna.lombardi@yahoo.com

tubicino sarà la causa della formazione di onde stazionarie all'interno della canna. Non resta che l'ultimo tocco, il più interessante: il cellulare. Nelle tasche di molti nostri studenti non è raro trovarlo.

Strumento di comunicazione, certo; strumento di calcolo, certo. Ora, però, è tempo di usarlo anche come strumento di misura. Il microfono dello smartphone preleverà il suono prodotto dalla Bic e una app gratuita (AndroSpectro Lite), che esegua il calcolo dello spettro del suono percepito, ci mostrerà le frequenze delle onde stazionarie nel tubo. Calcolo e misura affiancati in un oggetto di 120 g. Non è male. Cosa può desiderare di più un insegnante di matematica e fisica? Che ci piaccia o meno il laboratorio di fisica non è più della scuola, sta nelle mani dei nostri studenti e torna a casa con loro. A noi il compito di farglielo capire e introdurli all'utilizzo.

Ma torniamo all'esperimento: la classica biro trasparente, una volta privata del refill e chiuso il piccolo forellino che si trova sulla sua superficie laterale, si comporta approssimativamente come un cosiddetto risuonatore λ/4 (infatti il retro del tubo è chiuso dal tappino colorato). Soffiando all'imboccatura della biro si eccita l'aria nel tubo e si stimola la formazione di onde stazionarie al suo interno. Lo smartphone tramite microfono e algoritmo di calcolo della app ci mostra sul display lo spettrogramma del suono emesso dalla Bic. Analizzando i valori delle frequenze indicate sul display si può procedere al confronto con il model- lo teorico e trarre alcune conclusioni. (Il soffio deve essere leggero, all'imboccatura. I risultati mostrati in questo articolo sono stati ottenuti con un soffio perpendicolare all'asse longitudinale della penna, ma si può provare a ottenere misure analoghe anche soffiando lungo l'asse, cioè direttamente dentro il tubicino). In questi esperimenti abbiamo usato sistemi operativi Android e la app gratuita AndroSpectro Lite. È tuttavia possibile condurre analoghe esperienze con dispositivi Apple, per esempio, con la app SpectrumView.

### III. DATI E RISULTATO DELL'ESPERIMENTO

La Bic è lunga L = 0.133 m. Assumendo per il suono una velocità v = 340 m/s, la teoria dei tubi sonori chiusi ad una estremità prevede che si formino onde stazionarie con frequenze pari a:

$$f_n = \frac{(2n+1)v}{4L} = (2n+1) 6.39 \times 10^2 \; Hz. \tag{1}$$

Nella tabella sono raccolte le frequenze previste dalla teoria e quelle effettivamente misurate.

|       | Frequenze previste (Hz) | Frequenze misurate (Hz) |
|-------|-------------------------|-------------------------|
| $f_0$ | $6.39 \times 10^2$      | $5.78 \times 10^2$      |
| $f_1$ | $1.92 \times 10^3$      | $1.76 \times 10^3$      |
| $f_2$ | $3.20 \times 10^3$      | $3.18 \times 10^3$      |

*Tabella 1: frequenze teoriche e sperimentali penna Bic.*

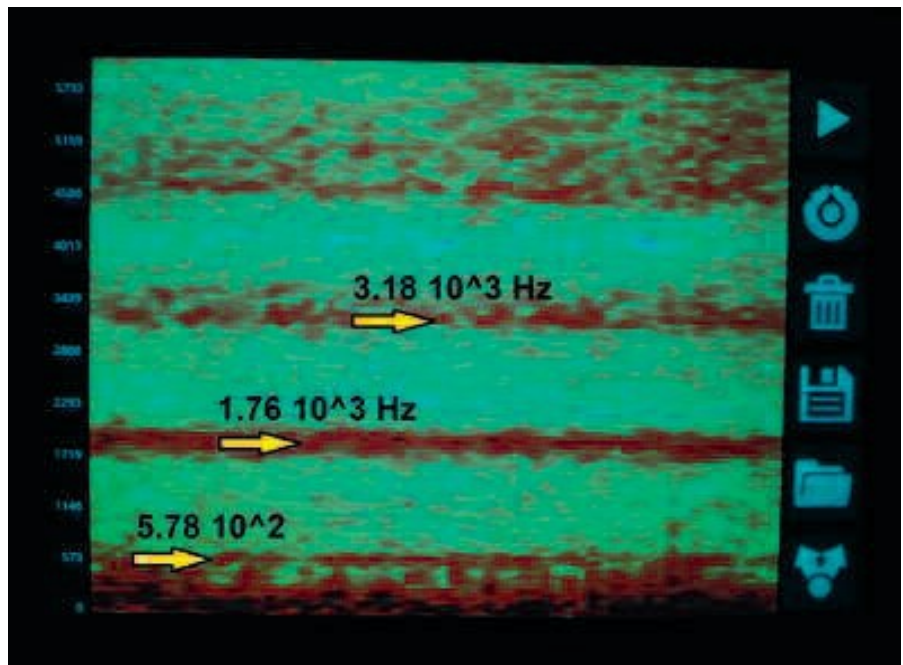

*Figura 1. Spettrogramma frequenze penna Bic.*

L'errore sperimentale, che va oltre quello dovuto ai fenomeni al bordo e che, come vedremo più avanti, prevede una apposita correzione, è a nostro parere da imputarsi in buona parte all'imboccatura della biro, che ha una forma conica e non perfettamente cilindrica. Questa imprecisione, tuttavia, non risulta inficiare lo scopo di questo stadio del lavoro che è quello, a partire dalla manipolazione di oggetti che i ragazzi hanno quotidianamente tra le mani, di osservare e misurare l'esistenza, all'interno di un tubo chiuso ad una estremità, di una ben precisa frequenza fondamentale e delle relative armoniche.

Per le misure delle frequenze, è possibile, poggiando il dito sul display e trascinando l'immagine dello spettrogramma in verticale, far coincidere una qualsiasi regione del grafico con una tacca numerata dell'asse verticale. In tal modo si possono ottenere valori numerici per le varie frequenze.

## IV. ACUSTICA DI UNA CANNUCCIA: FISICA AL BAR DELLA SCUOLA

Una volta accertato che in un tubo si formano onde di ben precise frequenze, tutte multiple di una frequenza fondamentale, possiamo passare ad uno stadio quantitativo, utilizzando un "tubo" più affidabile che troveremo al bar della scuola: una cannuccia. Per comodità studiamo il caso di tubo aperto ad entrambe le estremità. Procediamo come prima: soffiamo su un'estremità del tubo e sul display dello smartphone analizziamo le frequenze delle onde stazionarie.

Se la cannuccia non è troppo lunga, è possibile misurarne la lunghezza con un'altra applicazione gratuita, per esempio Smart Ruler Lite. Analogamente si può fare con il suo diametro così da ricavare il raggio; nel nostro caso $L = 0.175$ m e $R = 0.0025$ m.

Alla lunghezza reale della cannuccia è necessario aggiungere una correzione che tenga conto degli effetti ai bordi (0.6 R per ogni estremità aperta del tubo[i], nel nostro caso 2), per cui la lunghezza efficace del tubo risonante sarà:

$$L' = L + 2 \times 0.6\, R \qquad (2)$$

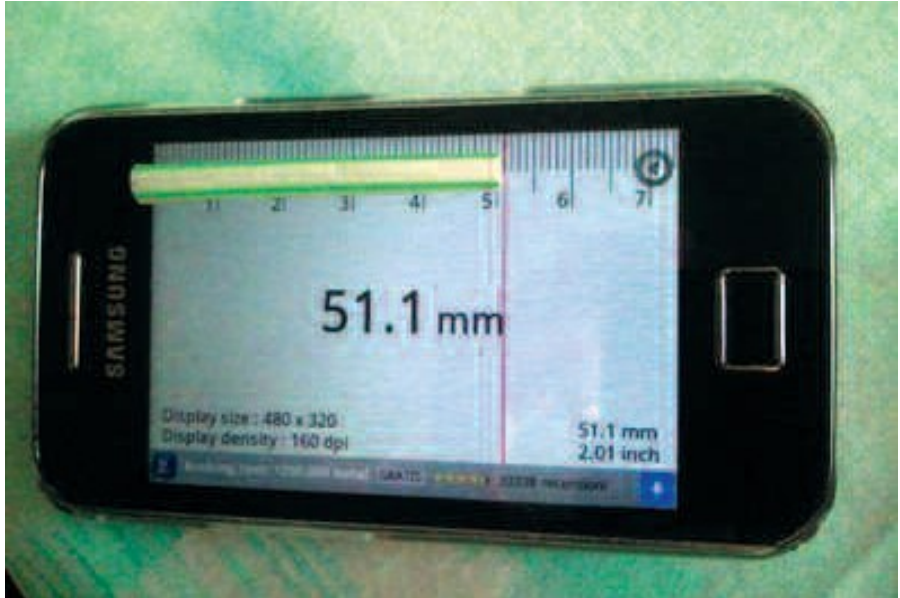

*Figura 2. La cannuccia viene misurata con una app del cellulare.*

Per la velocità del suono consideriamo ancora v = 340 m/s. Il modello teorico prevede:

$$f_n = \frac{(n+1)v}{L'} = (n+1)\, 9.7 \times 10^2\, Hz. \qquad (3)$$

|  | Frequenze previste (Hz) | Frequenze misurate (Hz) | Scarto percentuale dati sperim. - dati teorici |
|---|---|---|---|
| $f_0$ | $9.55 \times 10^2$ | $9.7 \times 10^2$ | 2% |
| $f_1$ | $1.91 \times 10^3$ | $1.95 \times 10^3$ | 2% |
| $f_2$ | $2.87 \times 10^3$ | $2.98 \times 10^3$ | 4% |
| $f_3$ | $3.82 \times 10^3$ | $3.97 \times 10^3$ | 4% |
| $f_4$ | $4.78 \times 10^3$ | $4.95 \times 10^3$ | 3% |

*Tabella 2: frequenze cannuccia aperta ad entrambe le estremità.*

Ecco come appare il display dello smartphone dopo il calcolo dello spettro del suono di una cannuccia aperta ad entrambe le estremità:

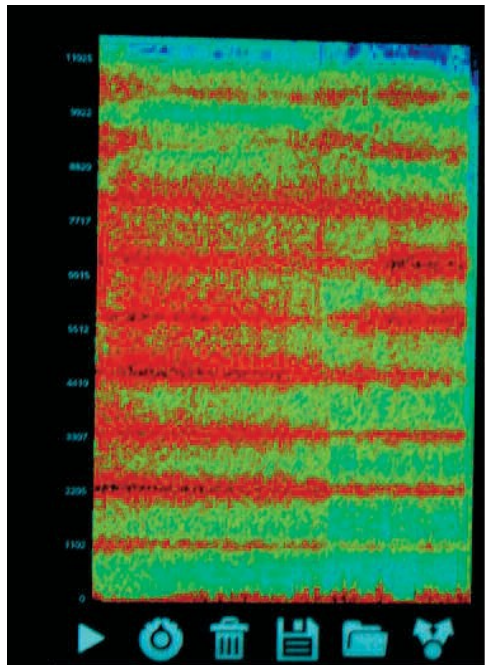

*Figura 3. Spettrogramma frequenze della cannuccia aperta ad entrambe le estremità.*

In verticale abbiamo le frequenze (in Hz) e in orizzontale il tempo. Le frequenze equispaziate sono ben visibili.

Per la cannuccia aperta ad una estremità e chiusa con un dito dall'altra abbiamo:

$$L' = L \times 0.6\,R \qquad (4)$$

e i valori attesi sono dati da:

$$f_n = \frac{2(n+1)v}{4L'} = (2n+1)\,4.9 \times 10^2\,Hz. \qquad (5)$$

|  | Frequenze previste (Hz) | Frequenze misurate (Hz) | Scarto percentuale dati sperim. - dati teorici |
|---|---|---|---|
| $f_0$ | $4.82 \times 10^2$ | $4.9 \times 10^2$ | 0,6% |
| $f_1$ | $1.44 \times 10^3$ | $1.45 \times 10^3$ | 0,7% |
| $f_2$ | $2.41 \times 10^3$ | $2.45 \times 10^3$ | 1% |
| $f_3$ | $3.37 \times 10^3$ | $3.40 \times 10^3$ | 0,9% |
| $f_4$ | $4.33 \times 10^3$ | $4.40 \times 10^3$ | 2% |

*Tabella 3: frequenze cannuccia chiusa ad una estremità.*

I ragazzi potranno verificare che, aprendo e chiudendo con un dito una apertura della cannuccia, le frequenze registrate dall'applicazione si modificano. In particolare, lasciando libere entrambe le aperture appaiono le frequenze intermedie rispetto a quelle emesse nel caso di una sola.

## V. CONSIDERAZIONI SULL'ERRORE NELLE MISURE CON LO SMARTPHONE

La app che calcola lo spettro del suono opera un campionamento a 22050 Hz, il numero di punti su cui lavora l'algoritmo che esegue la Fast Fourier Transform è pari a 1024. Ne consegue che la risoluzione in frequenza dell'apparato di misura[ii] è pari a:

$$df = \frac{frequenza\ di\ campionamento}{numero\ punti\ FFT} = 22\ Hz$$

Assumiamo allora che le misure in frequenza effettuate con lo smartphone abbiano come ultima cifra significativa (la prima affetta da errore) quella delle decine di hertz. A questo ci siamo attenuti nell'esporre i dati nelle tabelle soprastanti.

**Note**

i.   DUNCAN, J., STARLING, S.G. *A text book of physics*, MacMillan and Co, London, 1931, III – IV: Light and Sound, p. 421.

ii.  **http://www.vlf.it/fft_beginners/fft_beginners.html**